\documentclass[journal,twoside,web]{ieeecolor}
\usepackage{tmi}
\usepackage{cite}
\usepackage{amsmath,amssymb,amsfonts}
\usepackage{algorithm}
\usepackage{algorithmicx}
\usepackage[noend]{algpseudocode}
\usepackage{graphicx}
\usepackage{textcomp}
\usepackage{times}  
\usepackage{helvet} 
\usepackage{courier}  
\usepackage[hyphens]{url}  
\usepackage{mathrsfs}
\usepackage{multicol}
\usepackage{multirow}
\usepackage{setspace}
\usepackage{booktabs}
\usepackage{tabularx}
\usepackage{array}
\usepackage{balance}
\usepackage{xcolor}

\graphicspath{{./images/}}

\newcommand{\etal}{\emph{et~al. }}
\newcommand{\clA}{\mathcal{A}}

\newcommand{\clC}{\mathcal{C}}
\newcommand{\clE}{\mathcal{E}}
\newcommand{\clEh}{{\hat \clE}}
\newcommand{\clF}{\mathcal{F}}
\newcommand{\clH}{\mathcal{H}}

\newcommand{\clP}{\mathcal{P}}

\newcommand{\bbE}{\mathbb{E}}

\newcommand{\bfxx}{\mathbf{x}}

\newcolumntype{?}[1]{!{\vrule width #1}}

\makeatletter
\newcommand{\shorteq}{%
	\settowidth{\@tempdima}{-}
	\resizebox{\@tempdima}{\height}{=}%
}
\makeatother
\def\BibTeX{{\rm B\kern-.05em{\sc i\kern-.025em b}\kern-.08em
    T\kern-.1667em\lower.7ex\hbox{E}\kern-.125emX}}
\begin{document}
\title{AADG: Automatic Augmentation for Domain Generalization on Retinal Image Segmentation}
\author{Junyan Lyu, Yiqi Zhang, Yijin Huang, Li Lin, Pujin Cheng, Xiaoying Tang
\thanks{This study was supported by the Shenzhen Basic Research Program (JCYJ20190809120205578); the National Natural Science Foundation of China (62071210); the Shenzhen Basic Research Program (JCYJ20200925153847004); the High-level University Fund (G02236002). \textit{(Corresponding author: Xiaoying Tang)}}
\thanks{Junyan Lyu, Yiqi Zhang, Yijin Huang, Li Lin, Pujin Cheng and Xiaoying Tang are with the Department of Electronic and Electrical Engineering, Southern University of Science and Techonology, Shenzhen 518055, China (email: 12063003@mail.sustech.edu.cn; zhangyq2018@mail.sustech.edu.cn; 12150017@mail.sustech.edu.cn; 12150020@mail.sustech.edu.cn; 12032946@mail.sustech.edu.cn; tangxy@sustech.edu.cn).}
}

\maketitle

\begin{abstract}
Convolutional neural networks have been widely applied to medical image segmentation and have achieved considerable performance. However, the performance may be significantly affected by the domain gap between training data (source domain) and testing data (target domain). To address this issue, we propose a data manipulation based domain generalization method, called \emph{Automated Augmentation for Domain Generalization (AADG)}. Our \emph{AADG} framework can effectively sample data augmentation policies that generate novel domains and diversify the training set from an appropriate search space. Specifically, we introduce a novel proxy task maximizing the diversity among multiple augmented novel domains as measured by the Sinkhorn distance in a unit sphere space, making automated augmentation tractable. Adversarial training and deep reinforcement learning are employed to efficiently search the objectives. Quantitative and qualitative experiments on 11 publicly-accessible fundus image datasets (four for retinal vessel segmentation, four for optic disc and cup (OD/OC) segmentation and three for retinal lesion segmentation) are comprehensively performed. Two OCTA datasets for retinal vasculature segmentation are further involved to validate cross-modality generalization. Our proposed \emph{AADG} exhibits state-of-the-art generalization performance and outperforms existing approaches by considerable margins on retinal vessel, OD/OC and lesion segmentation tasks. The learned policies are empirically validated to be model-agnostic and can transfer well to other models. The source code is available at \text{https://github.com/CRazorback/AADG}.
\end{abstract}


\section{Introduction}
\label{sec:introduction}
\begin{figure}[t]
\begin{center}
    \includegraphics[width=0.95\linewidth]{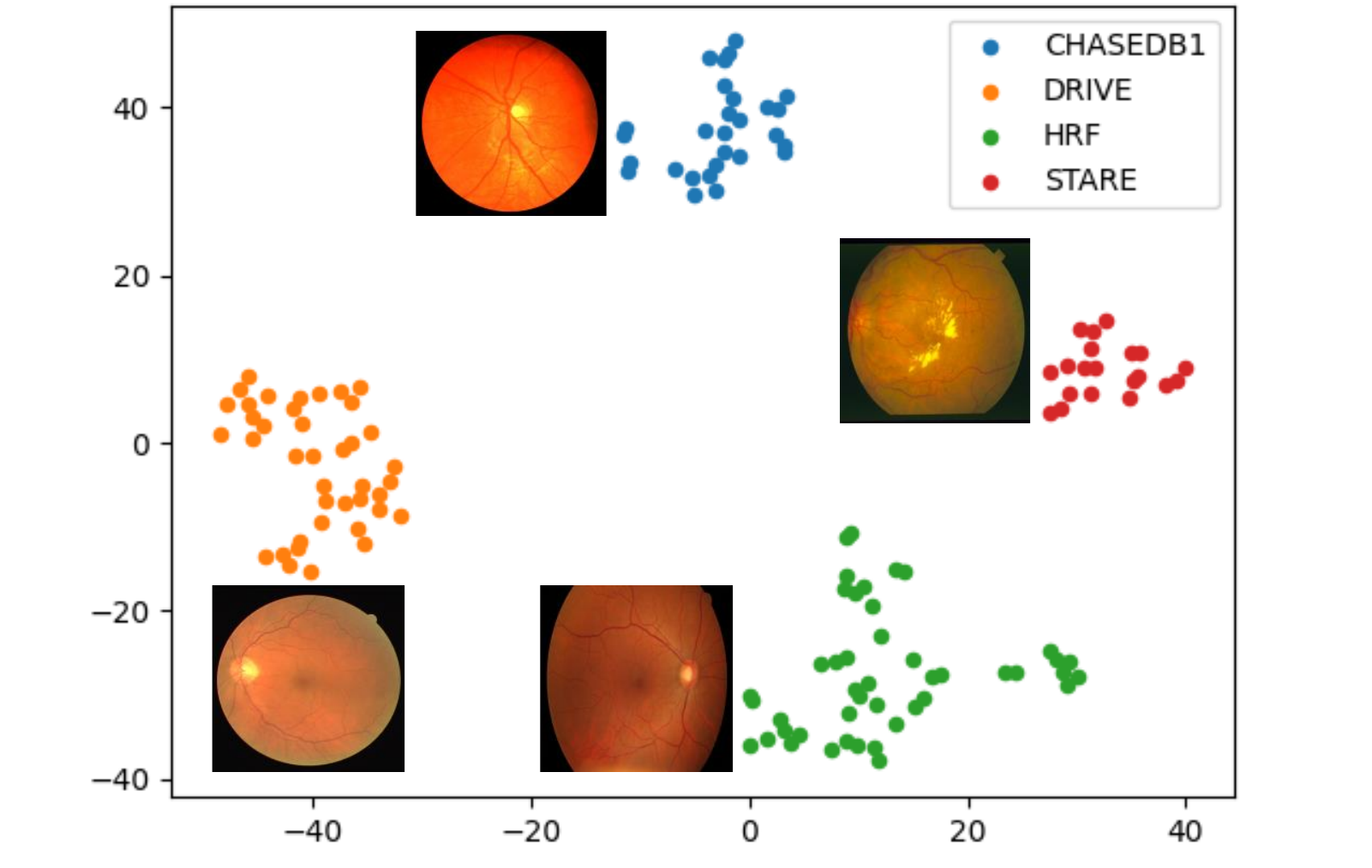}
\end{center}
\vspace{-3mm}
   \caption{t-SNE visualization of VGG16 features of fundus images acquired by different scanners and protocols.}
\label{fig:intro}
\end{figure}

\IEEEPARstart{R}{etinal} images play a vital role in diagnosing various eye diseases including diabetic retinopathy, glaucoma and age-related macular degeneration. The cup to disc ratio (CDR) is one of the most important biomarkers \cite{sivaswamy2015comprehensive} for glaucoma. To calculate CDR, segmentation of the optic disc (OD) and optic cup (OC) from retinal fundus images is a prerequisite. The retinal microvascular characteristics are associated with cerebral vascular and neural diseases \cite{debuc2017small,moss2015retinal}, wherein retinal vessel segmentation is essential. Recently, convolutional neural networks (CNNs) have been applied to automated OD/OC, retinal vessel and lesion segmentation, with superior performance having been obtained \cite{orlando2020refuge, lyu2019fundus, porwal2020idrid}. However, it is typically assumed that the training and testing images have similar distributions. Nevertheless, as shown in Fig.\ref{fig:intro}, the testing fundus images may be acquired by different scanners and protocols, resulting in great diversities in image appearance, contrast, quality and field of view. The performance of a well-trained CNN may deteriorate significantly on such a testing set \cite{zhang2020generalizing}. This largely discounts the practical value of CNNs in real-world applications.

To well accommodate images from different distributions, the best solution is to collect data from all possible domains to train a model. However, collecting and labeling medical image data from multiple sites is expensive and labor-intensive. Several techniques have been proposed to address this issue. Transfer learning is one of the most popular approaches to address the dataset bias and domain shift issues. It pretrains a model on a source domain, and finetunes the model on a different but related target domain. The major drawback of transfer learning is that it requires both images and labels from the target domain. Unsupervised domain adaptation (UDA) is proposed to alleviate the reliance on costly data annotation, but it still needs to access data from target domain. Furthermore, UDA needs to finetune the model to perform adaptation, which is often infeasible during clinical practice \cite{dou2019domain}. In such context, unseen domain generalization (DG) has become an active research direction and has gained broad interest and prospect in the medical image analysis community \cite{wang2021generalizing}. DG learns a universal model from diverse training datasets, which generalizes well to arbitrary unseen target domains without target data collection and finetuning. Existing DG methods generally focus on natural images. They can be divided into two categories, namely data manipulation based approaches and representation learning based approaches. Data manipulation based approaches focus on manipulating the input data to extend the training distribution and try to match the testing distribution. Towards this direction, data augmentation \cite{zhang2017mixup,volpi2018generalizing,yun2019cutmix} and data generation \cite{zhou2020learning,zhou2021domain} techniques have been explored to perform DG. Representation learning based approaches are further split into two lines. The first line focuses on learning domain-invariant representation \cite{li2018domain,fan2021adversarially}, which explicitly aligns features among domains in a latent space. The second line focuses on disentangling the latent features into a task-specified part and a domain-specified part for better generalization \cite{xu2014exploiting,nam2021reducing}.

A few attempts on DG have also been explored for medical image segmentation \cite{chen2020improving,zhang2020generalizing,wang2020dofe,liu2021feddg}. To be noted, traditional image transformations have shown their superiority in DG tasks such as segmenting medical images across different sites and different scanners. Compared with generative adversarial network (GAN) or gradient based image augmentation or generation, transformation based image augmentation has the unique advantage of preserving anatomical and structural information in the medical images of interest \cite{zhang2020generalizing}, which is crucial for medical image segmentation. Well-designed image transformation policies can generate novel domains and simulate sufficient imaging scenarios. However, manually designing a data augmentation policy requires adequate expertise and prior knowledge, especially in the medical image field. Moreover, the optimal policy may vary significantly across datasets and tasks. For instance, deforming during training is an effective data augmentation strategy for prostate segmentation, but not for atrial segmentation. To address this issue, automating the process of identifying an optimal data augmentation policy has recently emerged to improve the predictive performance on ImageNet \cite{cubuk2019autoaugment,lim2019fast} and COCO \cite{zoph2020learning}, and has been successfully applied to the medical segmentation decathlon challenge \cite{yang2019searching,xu2020automatic}. Yet, existing methods still assume that the training set has a similar distribution as the testing set. They simply maximize the accuracy on a validation set through a search algorithm. As such, current automated data augmentation methods are not applicable in DG scenarios. To bridge the gap, we propose to search for a domain generalized data augmentation policy on a novel proxy task.

In this article, we present a novel DG framework for retinal fundus images, namely \emph{Automated Augmentation for Domain Generalization (AADG)}. Instead of employing manually-designed data augmentation policies, \emph{AADG} regularizes network training by dynamically sampling data augmentation policies that can generate novel and diverse domains. The DG objective is achieved in two aspects: (i) the augmentation policy within an appropriate search space can generate novel domains and enlarge the distribution; and (ii) the policy controller maximizes the domain diversity across different augmented novel domains, as measured by the Sinkhorn distance. Consequently, the learned augmentation policy can help to train a model that generalizes well on different target domains. Our main contributions are summarized as follows:

\begin{itemize}
	\item  We propose a novel \emph{AADG} framework for DG on retinal fundus images. To the best of our knowledge, this is the first work to improve the generalization ability of retinal image segmentation models through automated augmentation.
	\item  We propose a novel proxy task for automated augmentation, which makes it feasible to search for the best data augmentation policies for DG. The proxy task helps the policy controller to efficiently sample a diverse training set.
	\item  We validate our proposed \emph{AADG} on fundus vessel (four datasets), OD/OC (four datasets), retinal lesion (three datasets) and OCTA vessel (two datasets) segmentation both qualitatively and quantitatively, exhibiting state-of-the-art (SOTA) performance and outperforming existing methods.
	\item  Extensive experiments show that the learned policies are explainable and can be easily transferred to other models without any modification, making it highly applicable.
\end{itemize}

\section{Related Work}
\subsection{Fundus Image Segmentation}  
Fundus image segmentation is a crucial step for quantitatively analyzing retinal structures such as vessels and OD/OC and diagnosing related diseases. In recent years, several CNN based approaches have been proposed and found to significantly outperform pervious approaches that rely on hand-crafted features. DRIU \cite{maninis2016deep} is the first method using CNN to understand fundus images, solving both retinal vessel segmentation and OD segmentation. Lyu \etal \cite{lyu2019fundus} proposed a novel fully convolutional network utilizing separable spatial and channel flow and densely adjacent vessel prediction to better capture spatial correlations between vessels. Fu \etal \cite{fu2018joint} performed joint OD/OC segmentation on polar coordinates using a multi-scale U-Net, which balanced the area proportion of OD and OC. CE-Net \cite{gu2019net} employed a novel dense atrous convolution block to further capture semantic information and preserve local information. Khojasteh \etal \cite{KHOJASTEH201962} explored and compared different deep learning techniques including ResNet, discriminative restricted Boltzmann machines and support vector machine to improve the performance of automatic exudate detection. However, most of these methods focused on improving the segmentation performance on a known dataset with better network architectures or utilizations of prior knowledge. They typically overfit a specific distribution and fail to generalize to unseen target datasets \cite{miyato2018virtual}.

\subsection{Automated Augmentation} 
Recently, several automated methods have been proposed to identify optimal data augmentation policies for specific datasets of interest. Most studies aim to improve the performance on image recognition tasks such as those related to ImageNet and CIFAR10. AutoAugment \cite{cubuk2019autoaugment} designed a search space and used deep reinforcement learning to search for the best policy yielding the highest validation accuracy on a target dataset. The search space has then been widely adopted by subsequent works. Fast AutoAugment \cite{lim2019fast} improved its predecessor via Bayesian optimization and efficient density matching, achieving comparable performance and significant accelerations. Adversarial AutoAugment \cite{zhang2019adversarial} simultaneously optimized the target task loss and a policy search loss, further reducing the error rate. Inspired by the success on natural images, a few research attempts have been made to conduct automated augmentation for medical images. Yang \etal \cite{yang2019searching} customized a search space including kernel filters, wherein the searching process was identical to that in AutoAugment. Xu \etal \cite{xu2020automatic} introduced differentiable data augmentation by means of stochastic relaxation, enabling gradient descent based optimization on both network parameters and augmentation parameters. The approaches proposed in those two works have both been validated on the 3D MRI segmentation tasks in the medical segmentation decathlon challenge. Yet, there has been no work applying automated augmentation to DG on medical datasets, especially on fundus image datasets.

\subsection{Domain Generalization on Natural Images}
DG has arisen to address the distribution shifts between a source domain and unseen target domains. Representation learning \cite{motiian2017unified,Li2018DomainGW,Peng2019DomainAL} largely contributes to the success of DG. For instance, Li \etal \cite{Li2018DomainGW} minimized the domain discrepancy among different source domains as measured by the maximum mean discrepancy (MMD) and successfully trained a domain-invariant autoencoder. Peng \etal \cite{Peng2019DomainAL} proposed a deep adversarial disentangled autoencoder, which explicitly disentangled the learned features into domain-specific ones and domain-invariant ones. There are also a few methods tackling DG by manipulating the input data. Zhou \etal \cite{zhou2020learning} imposed distribution divergence constraints on a GAN to generate pseudo-novel domains. Mixup \cite{zhang2017mixup,yun2019cutmix,verma2019manifold} is a popular data augmentation technique for DG \cite{Wang2020DomainMixLG,Zhou2021DomainGW,Qiao2021UncertaintyguidedMG}, which generates data by linear interpolations of image and the corresponding label pairs either in the input space or the latent space. Although these aforementioned studies were designed for natural images, they have also provided great insights for subsequent extensions to DG on medical images.

\subsection{Domain Generalization on Medical Images}
There has been substantial research interest in DG on medical images. Different from DG on natural images, traditional image transformations have shown great effectiveness in addressing domain shifts across different medical image datasets. Zhao \etal \cite{zhao2019data} leveraged shape and intensity information from unlabeled MRI images to deform and modify an atlas, providing significant improvements over existing works for one-shot brain MRI segmentation. Chen \etal \cite{chen2020improving} carefully designed data normalization and augmentation strategies to improve CNN-based model generalizability for DG on a cardiac MR image segmentation task. Ot{\'a}lora \etal \cite{otalora2019staining} also verified the superiority of traditional image transformations in computational pathology. BigAug \cite{zhang2020generalizing} generalized better to unseen domains by stacking more transformations. Liu \etal \cite{liu2021feddg} performed episodic learning in continuous frequency space (ELCFS), which smoothly bridged the distinct distributions of multiple source domains. Representation learning based methods have also been explored. Wang \etal \cite{wang2020dofe} presented a DoFE framework, embedding domain prior knowledge into the network of interest. DoFE employed an attention mechanism to combine memorized domain features with image features to perform generalizable fundus image segmentation. In light of these existing works, we further leverage traditional image transformations by automated augmentation techniques.

\section{Methodology}
\begin{figure*}[tbhp]
\begin{center}
    \includegraphics[width=0.95\textwidth]{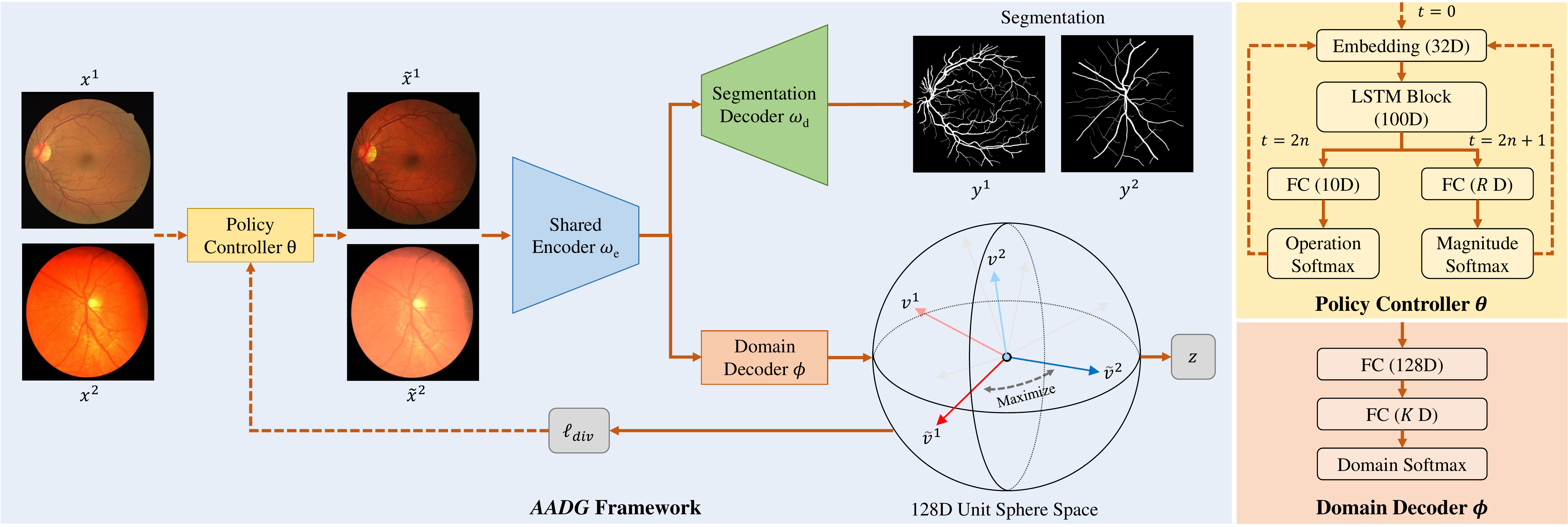}
\end{center}
   \caption{The overall training pipeline of \emph{AADG}. The image mini-batch is augmented by the policy controller $\theta$, and then sent into the shared encoder $\omega_\text{e}$ to extract semantic features. The semantic features are decoded by the segmentation decoder $\omega_\text{d}$ which predicts the segmentation. Meanwhile, the features are decoded by the domain decoder $\phi$ predicting the domain codes. $\ell_{div}$ measures the diversity of domains' latent distributions, which is maximized by DRL. Details of the policy controller and the domain decoder are illustrated in the right panel. The solid arrow indicates differentiable propagation and the dotted arrow means non-differentiable one.}
\label{fig:pipeline}
\end{figure*}

In this section, we firstly formulate the problem that \emph{AADG} addresses. Then, we introduce how we construct a search space and design the corresponding search objective. At last, we describe how we employ deep reinforcement learning (DRL) and adversarial training in \emph{AADG}. The overall framework is shown in Fig.~\ref{fig:pipeline}.

\subsection{Problem Formulation}
\label{sec:method}
Let $\mathcal{X}$ denote the input space and $\mathcal{Y}$ denote the output space. Each sample is denoted by $(\bfxx,y)\sim P$, where $P$ is the joint distribution of the input and the output. We assume that all possible target distributions follow an underlying hyper-distribution $P^{t}\sim \clP$. Then, the available source distributions from $K$ datasets also follow the same hyper-distribution $P^{1},P^{2},\cdots,P^{K} \sim \clP$. The goal of \emph{AADG} is to learn a classifier $h:\mathcal{X} \rightarrow \mathcal{Y}$ that minimizes the expected risk over the distribution of $\clP$, which can be described as
\begin{equation}
    \clE(h|\clP) := \bbE_{P \sim \clP} \bbE_{(\bfxx,y) \sim P} [\ell(h(\bfxx), y)],
\end{equation}
where $\ell$ measures the discrepancy on $\mathcal{Y}$. However, it is impossible to evaluate the expectation of the hyper-distribution since we cannot access data from all possible target domains. As such, we follow the empirical risk minimization (ERM) principle and estimate the expectation from available source domains, namely
\begin{equation}
    \clEh(h|\clP) := \frac{1}{K} \sum_{i=1}^K\frac{1}{N^i} \sum_{j=1}^{N^i} \ell(h(\bfxx^i_j), y^i_j),
\end{equation}
where $N^i$ denotes the number of training data from source domain $i$. 

Vapnik–Chervonenkis (VC) theorem has identified connections among the empirical risk, VC dimension, consistency and generalization \cite{vapnik2015uniform}. Considering a binary classifier $h\in \clH$ of a finite VC dimension $|\clH|_{VC}$, with a probability of no smaller than $1-\delta$, we have the following generalization bound
\begin{equation}
    \clE(h) - \clEh(h) \leqslant O\left(\left(\frac{|\mathcal{H}|_{\mathrm{VC}}-\log\delta}{\sum_{i=1}^{K} N^i}\right)^{\alpha}\right),
\label{upper_bound}
\end{equation}
where $\alpha \in [\frac{1}{2}, 1]$. According to this inequality, it is clear that the accuracy of estimating the expected risk can be improved by increasing the number of either the training samples or the source domains. Practically, data augmentation is an effective solution. We denote an optimal data augmentation function as $f \in \clF :\mathcal{X} \rightarrow \mathcal{X}$ which generates data $(\tilde{\bfxx},y)\sim P^{t}$. In other words, the function $f$ transforms an image from the source domain to a target domain in a label-preserving manner. Assume there is a finite number of $f$, we have 
\begin{equation}
    \clE(h|\clP) \leqslant O\left(\left(\frac{|\mathcal{H}|_{\mathrm{VC}}-\log\delta}{C\sum_{i=1}^{K} N^i}\right)^{\alpha}\right) + \clEh(h|\clP),
\label{upper_bound_augment}
\end{equation}
where $C$ is a finite constant that is proportional to the total number of novel domains. This clearly explains why traditional image transformations work better for DG on medical image segmentation. Traditional image transformations can simulate the intensity or geometry shifts caused by different scanner vendors, imaging protocols and conditions. As a result, they contribute to a larger $C$ and a lower expected risk. Nevertheless, some of the transformations can significantly change the image appearance and generate out-of-distribution (OOD) data. Traditional image transformations are defined as $\tilde{f} \in \clF :\mathcal{X} \rightarrow \mathcal{X}$ generating data $(\tilde{\bfxx},\tilde{y})\sim \tilde{\clP}$. Note that $\clP \subseteq \tilde{\clP}$ since $\tilde{\clP}$ includes OOD data. We suppose that
$$
\clE(h|\tilde{\clP}) - \clE(h|\clP) = \varepsilon_1 \geqslant 0,
$$
$$
\clEh(h|\tilde{\clP}) - \clEh(h|\clP) = \varepsilon_2 \geqslant 0.
$$
We replace $\clP$ in Eq.~\ref{upper_bound_augment} with $\tilde{\clP}$ since \emph{AADG} trains a classifier $h$ in $\tilde{\clP}$. Then, Eq.~\ref{upper_bound_augment} can be reformulated as
\begin{equation}
    \clE(h|\clP) \leqslant O\left(\left(\frac{|\mathcal{H}|_{\mathrm{VC}}-\log\delta}{C\sum_{i=1}^{K} N^i}\right)^{\alpha}\right) + \clEh(h|\clP) + \varepsilon_2 - \varepsilon_1,
\label{upper_bound_augment_final}
\end{equation}
where $\varepsilon_2 - \varepsilon_1$ is due to the distribution gap between $\clP$ and $\tilde{\clP}$. Eq.~\ref{upper_bound_augment_final} provides clear objectives for \emph{AADG} in three aspects:
\begin{enumerate}[]
    \item \textit{the total number of augmented novel domains $C$ being large,}
    \item \textit{the empirical risk $\clEh(h|\clP)$ being small,}
    \item \textit{the distribution gap between $\clP$ and $\tilde{\clP}$ being small,}
\end{enumerate}
wherein the second aspect can be satisfied by training a strong CNN on the source domain until convergence. In the following subsections, we focus on illustrating how we deal with the other two aspects via automated augmentation.



\subsection{Search Space}
We denote an image transformation operation as $o \in \mathcal{O}: \mathcal{X} \rightarrow \mathcal{X}$ defined on the input image space $\mathcal{X}$. Each operation $o$ has a magnitude $\lambda$. Each sub-policy $\tilde{f}$ consists of $L$ consecutive image transformation operations $\{o_{i}(x;\lambda_{i}),i=1,...,L\}$, which can be described as
\begin{equation}
    \tilde{f}=o_{1} \circ ... \circ o_{L} (x).
\end{equation}
Our final policy $F=\{\tilde{f}_1,\tilde{f}_2,...,\tilde{f}_S\}$ is a collection of $S$ sub-policies. The final policy randomly samples a sub-policy to perform a transformation. Theoretically, we can enumerate all operations of interest $o$ and their corresponding magnitudes $\lambda$ to get a large $C$. However, the convergence of $\clEh(h|\clP)$ cannot be guaranteed when trained on such a large dataset, even for SOTA segmentation models. Moreover, it will inevitably include a lot of OOD data, resulting in a large gap between $\clP$ and $\tilde{\clP}$. As such, automated augmentation is desired to identify a policy that can induce a large $C$ but a small distribution gap between $\clP$ and $\tilde{\clP}$.

The operations we search over include 10 color space transformations whose details are shown in Table~\ref{table:operations}. Spatial transformations such as shearing and translation are not taken into consideration, since they destroy the structural prior of both retinal vessels and OD/OC. To reduce variance during the search period, the magnitude of each operation is set to be within a certain range. Since the magnitudes of most operations are not differentiable, we discretize the range into $R$ values with uniform spacings. As a result, the search space of the policy has roughly $(10\times R)^{S\times L}$ possibilities, the optimization with respect to which can be solved by discrete search algorithms. Throughout this work, we consider different $R$ and $L$ to explore an appropriate search space.

\begin{table}[t]
\centering
\small
\caption{List of all image transformations that the controller could choose from during the search. Some transformations do not have the magnitude information (e.g. AutoContrast).}
\begin{tabular}{cc}
\toprule
  Operation Name & Range of Magnitudes \\ 
  \midrule
  AutoContrast & -\\ 
  Brightness & [0.1,1.9] \\  
  Color & [0.1,1.9]\\ 
  Contrast & [0.1,1.9]\\
  Cutout & [0,0.2] \\
  Equalize & -\\
  Invert & -\\ 
  Posterize & [4,8]\\
  Sharpness & [0.1,1.9] \\ 
  Solarize & [0,256] \\  
  
\bottomrule
\vspace{-5mm}
\end{tabular}
\label{table:operations}
\end{table}

\subsection{Search Objective}
In order to sample a policy that yields a large amount of novel domains $C$ from the search space, we need to maximize the difference among the augmented novel distributions
\begin{equation}
    \underset{F}{\text{ max }}  \ell_{div}=\displaystyle\sum_{i\neq j}^{K} d(F_a(X^{i}), F_b(X^{j})),
\label{eq:diversity}
\end{equation}
where $F_a$ and $F_b$ are different final policies. $X^{i},X^{j}$ are sampled mini-batches from different source domains in practice. In Eq.~\ref{eq:diversity}, $d(\cdot,\cdot)$ should be a distribution divergence measure in domain-specific latent or output spaces. Therefore, we introduce the domain classifier $c(\cdot, \phi) \in \clC$ to extract domain-specific features from the input images and predict the corresponding domain codes, where $\phi$ is the weight of the domain classifier. The domain classifier is optimized by minimizing the cross entropy loss
\begin{equation}
    \ell_{c}(\bfxx,z)=-z\log(c(\bfxx, \phi)),
\label{eq:advloss}
\end{equation}
where $z$ is the groundtruth domain code of $\bfxx$. With such a domain classifier, we can embed the domain information into an $n$ dimensional latent space.

Previous methods typically measure the probabilistic distribution divergence by KL-divergence \cite{Peng2019DomainAL,li2020maximum} or Wasserstein distance \cite{volpi2018generalizing,salimans2018improving,zhou2020learning}. We aim to maximize the divergence, but KL-divergence is numerically unstable when its value is large. Thus, Wasserstein distance is adopted in our study, defined as
\begin{equation}
d_{ot}(\tilde{\clP}^i, \tilde{\clP}^j) = \inf_{\pi \in \Pi(\tilde{\clP}^i, \tilde{\clP}^j)} \mathbb{E}_{\bfxx_i, \bfxx_j \sim \pi} [\ell(\bfxx_i, \bfxx_j)],
\label{eq:wasserstein}
\end{equation}
where $\Pi(\tilde{\clP}^i, \tilde{\clP}^j)$ is the set of joint distributions $\pi(\bfxx_i, \bfxx_j)$. $\ell(\bfxx_i, \bfxx_j)$ is a transport cost function which we take to be the cosine similarity since it is more statistically efficient than the Euclidean distance in high dimensional settings \cite{salimans2018improving}. 

Given that the cost of computing optimal transport distances grows exponentially, we adopt an entropically smoothed generalization of Wasserstein distance, namely Sinkhorn distance \cite{cuturi2013sinkhorn}
\begin{equation}
    d_{sinkhorn}(\tilde{\clP}^i, \tilde{\clP}^j) = \inf_{M \in \mathcal{M}} \text{Tr}[MC^{\text{T}}],
\label{eq:sinkhorn}
\end{equation}
where $M$ is the soft matching matrix replacing the coupling distribution $\pi$ and $C$ is the pair-wise cosine distance matrix computed over two sets. To be noted, minimization over the soft matching matrix $M$ can be efficiently solved on GPU using the Sinkhorn algorithm, which makes Eq.~\ref{eq:diversity} tractable. $d_{sinkhorn}(\cdot,\cdot)$ does not need to be valid for probability distributions, since we use it in a latent space instead of the output space.

Apart from increasing the diversity of the augmented novel distributions, Eq.~\ref{eq:diversity} also implicitly avoids OOD data. There is no doubt that image transformations with extreme values of magnitudes may severely distort images. Intuitively, adjusting the brightness of an image with a magnitude of 0 gives a completely black image, whereas a magnitude of 1 gives a completely white one. They are both indistinguishable images in terms of domains. According to the weaker invariance assumption \cite{ratner2017learning}, these OOD data will be mapped into a "null" class but not other domains. This induces a low diversity as measured by Eq.~\ref{eq:sinkhorn} in a unit sphere space. Consequently, the distribution gap between $\clP$ and $\tilde{\clP}$ can be bridged by solving Eq.~\ref{eq:diversity}. So far, we have met all the objectives revealed by Eq.~\ref{upper_bound_augment_final}.

\subsection{Adversarial Training}
Searching an augmentation policy is a bi-level optimization problem, formulated as follows
\begin{equation}
    \underset{F}{\text{ max }}\ell_{div}, \quad s.t. \quad \phi^* = \underset{\phi}{\text{argmin }} \ell_{c}.
\end{equation}
We denote $h(\cdot,\omega)\in \clH$ as the segmentation model, where $\omega$ is the weight of the segmentation model and is obtained by minimizing the cross entropy loss
\begin{equation}
    \ell_{h}(\bfxx,y)=-y\log(h(\bfxx, \omega)).
\label{eq:segloss}
\end{equation}
$y$ is the groundtruth segmentation of $\bfxx$. We can then train a domain generalized segmentation model with the best policy $F$. Previous studies \cite{zhang2019adversarial,liu2021divaug} have found that jointly optimizing the segmentation model training and the augmentation policy searching achieved better performance with higher computational efficiency. We further fit the original optimization problem into an adversarial learning framework, which jointly optimizes the segmentation model, domain classifier and policy controller via
\begin{equation}
     \underset{\omega,\phi}{\text{min}}\underset{F}{\text{ max }}\ell_{h}+\ell_{c}+\ell_{div}.
\label{eq:8}
\end{equation}
Since the segmentation model and the domain classifier are both consistently updated by gradient descent, the feature averaging effect of augmentation \cite{dao2019kernel} tends to increase the transformation invariance of the model, and thus the optimal data augmentation policy changes throughout the optimization process. That is to say, adversarial learning encourages the policy controller to sample more diverse sub-policies and avoid local maxima. Consequently, the search space of the policy is significantly enlarged to $(10\times R)^{S\times L\times E}$, where $E$ denotes the times the policy controller updates for.

In order to solve this problem, DRL is adopted as our discrete search algorithm for its simplicity. DRL contains an augmentation policy controller based on recurrent neural network (RNN), an environment and a training algorithm. As shown in Fig.~\ref{fig:pipeline}, the RNN of the policy controller is implemented by a single hidden layer LSTM with 100 hidden units. The prediction layers of the LSTM include two fully-connected layers which respectively provide softmax predictions of the operation types and the magnitude levels. At each step, the controller produces a softmax prediction and then embeds it as the input of the next step, with an embedding size of 32. The total step it takes the controller to predict a sub-policy $\tilde{f}$ is $2\times L$, producing $2\times L$ softmax predictions of the operation types and magnitudes. The controller predicts a policy $F$ by repeating the aforementioned process for $S$ times. An augmentation policy controller $a(\theta) \in \clA$ is optimized based on its policy gradients
\begin{equation}
    \triangledown_{\theta}J(\theta)=\triangledown_{\theta}\log a(\theta)\ell_{div},
\label{eq:policyloss}
\end{equation}
where $\ell_{div}$ evaluates the policy prediction $a(\theta)$ and is regarded as the reward signal.

To stabilize optimization, we normalize $\triangledown_{\theta}J(\theta)$ among $B$ sampling policies. In this study, $B$ is empirically set to be six. All steps of training \emph{AADG} are described in Algorithm~\ref{algo:1}. 

\begin{algorithm}[t]
\setstretch{1.1}
	\caption{Adversarial Training of Automated Augmentation for Domain Generalization}
	\label{algo:1}
	{\bf Initialization:} segmentation model $h(\cdot, \omega)$, domain classifier $c(\cdot, \phi)$, augmentation policy controller $a(\theta)$\\
	{\bf Input:} source images $\bfxx$, segmentation labels $y$, domain codes $z$ \\
	{\bf Output:} $\omega^{*}$, $\phi^{*}$, $\theta^{*}$
	\begin{algorithmic}[1]
	\For{$1 \leq e \leq E$ }
	      \State Initialize $\ell_{div}^{b} = 0,\forall b\in\{1,2,...,B\}$;
	      \State Sample $B$ augmentation policies;
		  \For{$ 1 \leq t \leq T$}
		  	\State Augment the image mini-batch with $B$ policies;
		  	\State Update $\phi$ by minimizing $\ell_c$;
		  	\State Update $\omega$ by minimizing $\ell_h$;
		  	\State Update $\ell_{div}$ via moving average;
		  \EndFor
		  \State Collect $\{\ell_{div}^1, \ell_{div}^2,...,\ell_{div}^B\}$;
		  \State Normalize $\ell_{div}^b$ among $B$ instances;
		  \State Update $\theta$ by maximizing $\triangledown_{\theta}J(\theta)$;
	\EndFor
	\end{algorithmic}
\end{algorithm}

\section{Experiments}

\subsection{Datasets}
We comprehensively evaluate \emph{AADG} on retinal vessel segmentation based on retinal fundus images. Specifically, we conduct experiments on four publicly-available datasets including STARE\cite{hoover2000locating}, HRF\cite{budai2013robust}, DRIVE\cite{staal2004ridge} and CHASEDB1\cite{fraz2012ensemble}. The sample size is respective 20, 45, 40 and 28. To evaluate \emph{AADG}'s cross-modality domain generalization performance, we also include two OCTA datasets during inference, namely the OCTA-500 \cite{li2020ipn} and ROSE \cite{ma2020rose}. The sample size is respective 445 and 39. We further validate the effectiveness of \emph{AADG} on OD/OC segmentation based on retinal fundus images from Drishti-GS \cite{sivaswamy2015comprehensive}, RIM-ONE-r3 \cite{fumero2011rim}, REFUGE-train and REFUGE-val \cite{orlando2020refuge}. The sample size is respective 101, 159, 400 and 400. For both tasks, the images are acquired from different clinical centers, scanners and populations. The training and validation sets are divided in the same way as those in previous works \cite{wang2020dofe, liu2021feddg, peng2022student}. 

Further on we evaluate \emph{AADG} for lesion segmentation on three publicly-available datasets of IDRiD \cite{ting2016diabetic}, E-Ophtha \cite{decenciere2013teleophta} and DDR \cite{li2019diagnostic}. Only annotations of the hard exudates (HEs) are commonly available in all three datasets, and thus images with HEs are used in this experiment. The sample size is respective 81, 82 and 757. We adopt the official splits of IDRiD and DDR for training and testing. The images from E-Ophtha are randomly divided into 50/32 images for training/testing.

\subsection{Implementation Details}
\emph{AADG} is implemented by PyTorch \cite{paszke2019pytorch} on a workstation equipped with NVIDIA Tesla A100. We adopt DeepLabv3+ with the backbone of MobileNetv2 as our segmentation model so as to fairly compare with other SOTA methods. The backbone is initialized with ImageNet weights. The domain classifier shares a feature encoder with the segmentation model, and has a simple yet effective domain encoder, as illustrated in Fig.~\ref{fig:pipeline}. Adam optimizers are used for the segmentation model and the domain classifier with learning rates of 0.001, and also for the policy controller with a learning rate of 0.00035. The batch size is set to be 24. We use the Proximal Policy Optimization (PPO) algorithm to train the policy controller. We apply a tanh constant of 2.5 and a softmax temperature of 2 to the controller's logits\cite{bello2016neural}, and further add an entropy penalty with a weight of 0.00001 to regularize the controller. By default, $R=10,S=5,L=2$. For OD/OC segmentation, $E=150$. For retinal vessel and lesion segmentation, $E=300$ due to the small sample sizes.

For OD/OC segmentation, we first crop images centering OD and then resize to be $256\times 256$ in the pre-processing stage. For vessel and lesion segmentation, the images are resized to be $512\times 512$. For all tasks, we randomly scale, crop, rotate and flip images during training to expand the training sets. In other words, those four types of operations (scaling, cropping, rotation, and flipping) are treated as our default training set augmentation techniques.

The performance of \emph{AADG} and other compared SOTA methods is evaluated using the Dice Similarity Coefficient (DSC) on all the segmentation tasks. Accuracy (ACC) and the area under the receiving operator characteristic curve (AUC-ROC) are further adopted as the evaluation metrics for the vessel segmentation task on fundus images. To ensure the reliability and robustness of our experimental results, we repeat each experiment for five times and report the mean value.

\begin{table*}[h!]
\centering
\caption{{Quantitative comparisons with SOTA DG methods for fundus vessel segmentation. The results of domain A are obtained from the model trained with images from the other domains. Domains A, B, C, D are respective STARE, HRF, DRIVE and CHASEDB1. Top 1 results are highlighted in \textbf{bold}. '-' indicates that DoFE does not report DSC values.}}
\label{tab:vesselsota}
\renewcommand{\arraystretch}{1.25}
\resizebox{\textwidth}{!}{
\begin{tabular}{l|llll|l|llll|l|llll|l}
\toprule
\multicolumn{1}{c|}{\multirow{2}{*}{\textbf{Method}}} & \multicolumn{4}{c|}{\textbf{DSC} $\uparrow$} &                       \multicolumn{1}{c|}{\multirow{2}{*}{\textbf{Average}}} & \multicolumn{4}{c|}{\textbf{AUC-ROC} $ \uparrow$} & \multicolumn{1}{c|}{\multirow{2}{*}{\textbf{Average}}} & \multicolumn{4}{c|}{\textbf{ACC} $ \uparrow$} & \multicolumn{1}{c}{\multirow{2}{*}{\textbf{Average}}}\\ \cline{2-5} \cline{7-10} \cline{12-15}
\multicolumn{1}{c|}{}    & \multicolumn{1}{c}{\textbf{A}} & \multicolumn{1}{c}{\textbf{B}} & \multicolumn{1}{c}{\textbf{C}} & \multicolumn{1}{c|}{\textbf{D}} &  & \multicolumn{1}{c}{\textbf{A}} & \multicolumn{1}{c}{\textbf{B}} & \multicolumn{1}{c}{\textbf{C}} & \multicolumn{1}{c|}{\textbf{D}} &  & \multicolumn{1}{c}{\textbf{A}} & \multicolumn{1}{c}{\textbf{B}} & \multicolumn{1}{c}{\textbf{C}} & \multicolumn{1}{c|}{\textbf{D}}               \\ \hline
Baseline & 76.32 & 72.23 & 76.27 & 75.71 & \multicolumn{1}{c|}{75.13} & 97.36 & 92.18 & 94.50 & 95.97 & \multicolumn{1}{c|}{95.01} & 94.10 & 90.46 & 92.07 & 93.45 & \multicolumn{1}{c}{92.52}\\ \hline
19'M-Mixup\cite{verma2019manifold} & 61.89 & 60.41 & 58.22 & 63.21 & \multicolumn{1}{c|}{60.93} & 96.02 & 91.76 & 94.30 & 95.45 & \multicolumn{1}{c|}{94.38} & 92.32 & 89.25 & 89.05 & 92.46 & \multicolumn{1}{c}{90.77}\\ 
19'CutMix\cite{yun2019cutmix} & 74.71 & 69.86 & 74.34 & 75.08 & \multicolumn{1}{c|}{73.50} & 97.46 & 92.06 & 94.44 & 95.70 & \multicolumn{1}{c|}{94.92} & 94.03 & 90.71 & 91.84 & 93.80 & \multicolumn{1}{c}{92.60}\\ \hline
20'BigAug \cite{zhang2020generalizing} & 79.61 & 70.06 & 76.42 & 76.50 & \multicolumn{1}{c|}{75.65} & 97.36 & 90.59 & 94.78 & 96.09 & \multicolumn{1}{c|}{94.70} & 94.21 & 89.71 & 91.96 & 93.59 & \multicolumn{1}{c}{92.37}\\
20'DoFE \cite{wang2020dofe} & \multicolumn{1}{c}{-} & \multicolumn{1}{c}{-} & \multicolumn{1}{c}{-} & \multicolumn{1}{c|}{-} & \multicolumn{1}{c|}{-} & 97.25 & 89.48 & 94.23 & 96.28 & \multicolumn{1}{c|}{94.31} & 94.57 & 89.49 &  94.11 & 90.51 & \multicolumn{1}{c}{92.17}\\ 
21'ELCFS \cite{liu2021feddg} & 80.92 & 71.85 & 76.61 & 76.40 & \multicolumn{1}{c|}{76.44} & 97.82 & \textbf{92.18} & 95.14 & 96.29 & \multicolumn{1}{c|}{95.36} & 94.54 & \textbf{90.57} & 92.27 & 93.44 & \multicolumn{1}{c}{92.71}\\
\hline
\textbf{\emph{AADG}} & \textbf{81.79} & \textbf{72.57} & \textbf{77.70} & \textbf{78.34} & \multicolumn{1}{c|}{\textbf{77.60}} & \textbf{97.96} & 91.93 & \textbf{95.21} & \textbf{96.95} & \multicolumn{1}{c|}{\textbf{95.52}}& \textbf{94.75} & 90.49 & \textbf{92.41} & \textbf{93.84} & \multicolumn{1}{c}{\textbf{92.87}}\\ 
\bottomrule
\end{tabular}
}
\end{table*}

\begin{table*}[h!]
\centering
\caption{Quantitative DSC comparisons with SOTA DG methods on OD/OC segmentation. Domains A, B, C, D are respective Drishiti-GS, RIM-ONE-r3, REFUGE-train and REFUGE-val. Top 1 results are highlighted in \textbf{bold}.}
\label{tab:odocsota}
\renewcommand{\arraystretch}{1.2}
\resizebox{0.8\textwidth}{!}{
\begin{tabular}{l|lllll|lllll|l}
\toprule
\multicolumn{1}{c|}{\multirow{2}{*}{\textbf{Method}}} & \multicolumn{5}{c|}{\textbf{OD Segmentation}}                                                                         & \multicolumn{5}{c|}{\textbf{OC Segmentation}}           & \multicolumn{1}{c}{\multirow{2}{*}{\textbf{Average}}} \\ \cline{2-11}
\multicolumn{1}{c|}{}    & \multicolumn{1}{c}{\textbf{A}} & \multicolumn{1}{c}{\textbf{B}} & \multicolumn{1}{c}{\textbf{C}} & \multicolumn{1}{c|}{\textbf{D}}& \multicolumn{1}{c|}{\textbf{Average}} & \multicolumn{1}{c}{\textbf{A}} & \multicolumn{1}{c}{\textbf{B}} & \multicolumn{1}{c}{\textbf{C}} & \multicolumn{1}{c|}{\textbf{D}} &  \multicolumn{1}{c|}{\textbf{Average}} &              \\ \hline
Baseline & 94.96 & 89.69 & 89.33 & 90.09 & \multicolumn{1}{|c|}{91.02} & 77.03 & 78.21 & 80.28 & 84.74 & \multicolumn{1}{|c|}{80.07} & \multicolumn{1}{c}{85.54} \\ \hline
19'M-Mixup \cite{verma2019manifold} & 94.48 & 89.38 & 92.17 & 90.82 & \multicolumn{1}{|c|}{91.71} & 79.27 & 75.41 & 83.01 & 86.73 & \multicolumn{1}{|c|}{81.11} & \multicolumn{1}{c}{86.41} \\
19'CutMix \cite{yun2019cutmix} & 93.83 & 91.97 & 90.13 & 88.79 & \multicolumn{1}{|c|}{91.18} & 76.97 & 81.02 & 83.42 & 86.83 & \multicolumn{1}{|c|}{82.46} & \multicolumn{1}{c}{86.62} \\ \hline
20'BigAug \cite{zhang2020generalizing} & 92.20 & 90.77 & \textbf{94.02} & 90.66 & \multicolumn{1}{|c|}{91.91} & 75.63 & 80.80 & 84.32 & 86.24 & \multicolumn{1}{|c|}{81.75} & \multicolumn{1}{c}{86.83} \\ 
20'DoFE \cite{wang2020dofe} & \textbf{95.59} & 89.37 & 91.98 & 93.32 & \multicolumn{1}{|c|}{92.57} & 83.59 & 80.00 & 86.66 & 87.04 & \multicolumn{1}{|c|}{84.31} & \multicolumn{1}{c}{88.44} \\ 
21'ELCFS \cite{liu2021feddg} & 95.37 & 87.52 & 93.37 & \textbf{94.50} & \multicolumn{1}{|c|}{92.69} & 84.13 & 71.88 & 83.94 & 85.51 & \multicolumn{1}{|c|}{81.37} & \multicolumn{1}{c}{87.03} \\
\hline
\textbf{\emph{AADG}} & 95.47 & \textbf{92.28} & 93.55 & 92.55 & \multicolumn{1}{|c|}{\textbf{93.46}} & \textbf{85.93} & \textbf{81.24} & \textbf{86.94} & \textbf{87.86} & \multicolumn{1}{|c|}{\textbf{85.50}} & \multicolumn{1}{c}{\textbf{89.48}} \\ 
\bottomrule
\end{tabular}
}
\end{table*}

\begin{table}[t]
\caption{Quantitative DSC comparisons with SOTA DG methods on lesion segmentation. Domains A, B, C are respective IDRiD, E-ophtha and DDR. Top 1 results are highlighted in \textbf{bold}.}
\centering
\renewcommand{\arraystretch}{1.2}
\resizebox{0.72\columnwidth}{!}{
\begin{tabular}{l|lll|l}
\toprule
\multicolumn{1}{c|}{\textbf{Method}} & \multicolumn{1}{c}{\textbf{A}} & \multicolumn{1}{c}{\textbf{B}} & \multicolumn{1}{c|}{\textbf{C}} &     \multicolumn{1}{c}{\textbf{Average}}                        \\ \hline
Baseline & 53.94 & 50.77 & 32.94  & \multicolumn{1}{c}{45.88} \\
21'ELCFS & 56.90 & 52.29 & 34.27 & \multicolumn{1}{c}{47.82}\\
\hline
\emph{AADG} & \textbf{59.22} & \textbf{53.51} & \textbf{36.58} & \multicolumn{1}{c}{\textbf{49.77}}\\
\bottomrule
\end{tabular}
}
\label{tab:hardex}
\end{table}

\begin{table}[t]
\caption{Quantitative DSC comparisons with SOTA DG methods on OCTA vessel segmentation. Domains A, B are respective OCTA-500 and ROSE. Top 1 results are highlighted in \textbf{bold}.}
\centering
\renewcommand{\arraystretch}{1.2}
\resizebox{0.58\columnwidth}{!}{
\begin{tabular}{l|ll|l}
\toprule
\multicolumn{1}{c|}{\textbf{Method}} & \multicolumn{1}{c}{\textbf{A}} & \multicolumn{1}{c|}{\textbf{B}} &     \multicolumn{1}{c}{\textbf{Average}}                        \\ \hline
Baseline & 3.81 & 1.69 & \multicolumn{1}{c}{2.75} \\
21'ELCFS & 8.51 & 6.13  & \multicolumn{1}{c}{7.32}\\
\hline
\emph{AADG} & \textbf{61.57} & \textbf{50.78} & \multicolumn{1}{c}{\textbf{56.18}}\\
\bottomrule
\end{tabular}
}
\label{tab:octa}
\end{table}

\subsection{Comparisons with SOTA}
In our experiments, we follow previous literature and use the leave-one-domain-out strategy to validate the effectiveness of all DG methods of interest. Specifically, we train a model on source domains A, B, C and test it on the left-out unseen target domain D. The baseline is trained on the union of all source domains without any generalization technique, which simply concatenates multiple source domains. Please note that the baseline is equipped with the backbone of MobileNetv2 if not specified the backbone. Table~\ref{tab:vesselsota}, Table~\ref{tab:odocsota}, Table~\ref{tab:hardex} and Table~\ref{tab:octa} present quantitative comparisons between the baseline and \emph{AADG}. The proposed \emph{AADG} outperforms the baseline by a large margin (2.47\% increase for fundus vessel segmentation, 3.94\% increase for OD/OC segmentation, 3.89\% increase for retinal lesion segmentation and 53.43\% increase for OCTA vessel segmentation in terms of the average DSC). These results clearly identify the superior generalization ability of \emph{AADG}.

\begin{figure*}[thbp]
\begin{center}
    \includegraphics[width=0.88\textwidth]{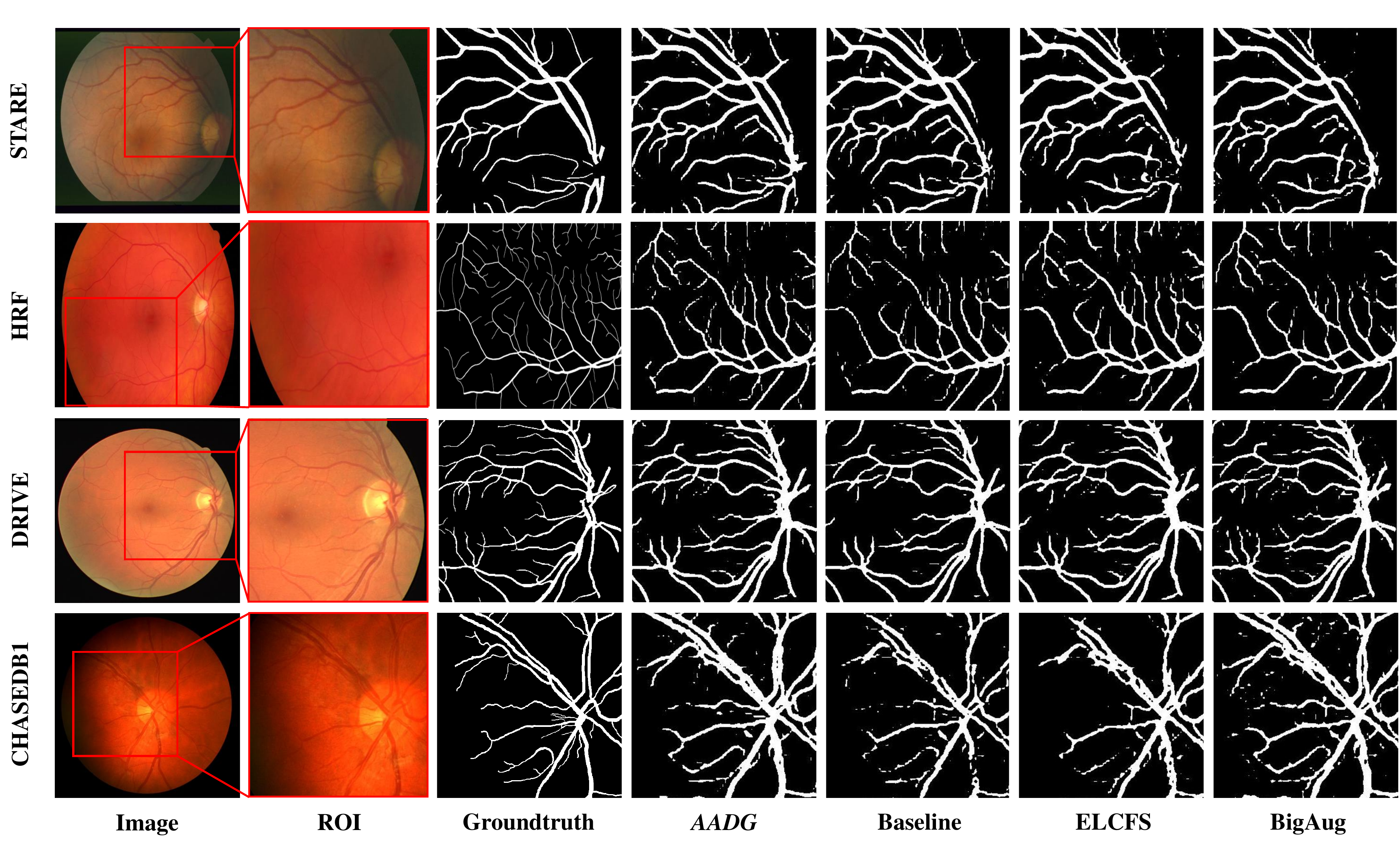}
\end{center}
   \caption{Qualitative comparisons of \emph{AADG} and existing SOTA DG methods on fundus vessel segmentation.}
\label{fig:rvs}
\end{figure*}

\begin{figure*}[thbp]
\begin{center}
    \includegraphics[width=0.88\textwidth]{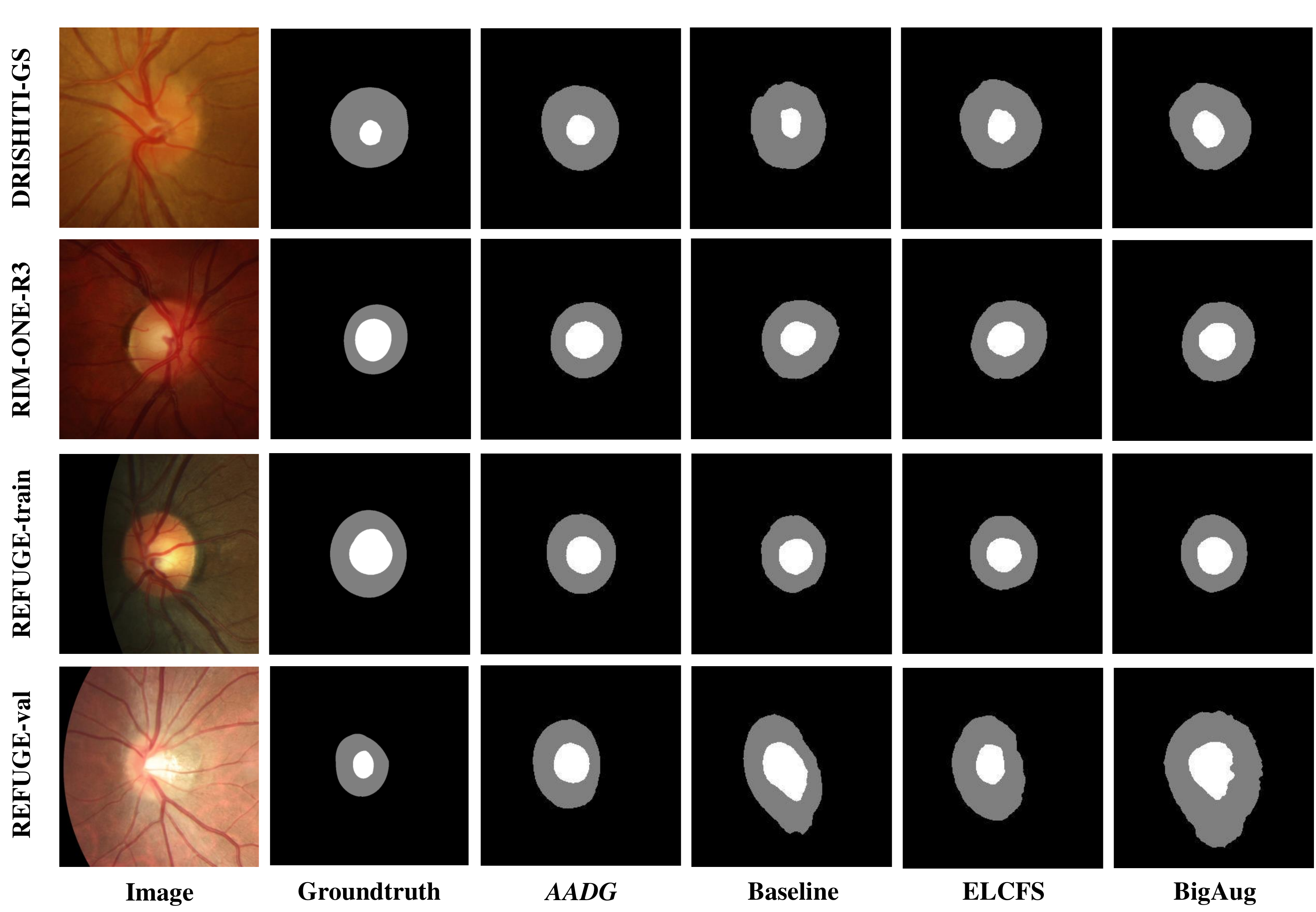}
\end{center}
   \caption{Qualitative comparisons of \emph{AADG} and existing SOTA DG methods on OD/OC segmentation.}
\label{fig:optic}
\end{figure*}

\subsubsection{Comparisons with DG methods for natural images} In Tables~\ref{tab:vesselsota} and~\ref{tab:odocsota}, we also compare \emph{AADG} with two DG methods that are commonly used for natural images, namely manifold mixup (M-mixup) and CutMix. M-mixup achieves superior segmentation performance over baseline for OD/OC, but inferior segmentation performance for retinal vessel. This is because retinal vessel is more sophisticated in geometry than OD/OC, for which M-mixup may greatly distort the task-specific latent features. CutMix performs mixup in the image space, inducing a slight decline in DSC but a mild increase in ACC for vessel segmentation. These results empirically explain why DG methods for natural images are rarely applied to medical images.

\subsubsection{Comparisons with DG methods for medical images} We then compare \emph{AADG} with three DG methods for medical images, including DoFE, BigAug and ELCFS, also in Tables~\ref{tab:vesselsota} and~\ref{tab:odocsota}. DoFE aggregates domain information in a feature space, boosting the generalization performance for segmenting OD/OC with an increase of 2.90\% in DSC over the baseline. However, DoFE fails to generalize to segment retinal vessels. BigAug applies a stack of image transformations with handcrafted parameters, exhibiting consistent improvements over the baseline for both tasks. Yet, BigAug conservatively designs the magnitude ranges (e.g. adjusting the brightness by randomly shifting the intensity with a magnitude ranging between [-0.1, 0.1]). It generates novel domains but of insufficient diversity. Moreover, BigAug includes spatial variations such as deformation, which is inapplicable for vessel segmentation. These factors result in consistent but marginal improvements (0.52\% and 1.29\% increases in DSC for the two tasks). The comparison between BigAug and \emph{AADG} reveals that handcrafted data augmentation can be improved by automated augmentation. ELCFS improves the baseline by an increase of 1.31\% for vessel segmentation and an increase of 1.49\% for OD/OC segmentation in DSC. Such improvements benefit from the frequency space interpolation mechanism, which can continuously generate images exhibiting characteristics of other domains. Our \emph{AADG} further improves ELCFS, enhancing the segmentation performance in terms of all metrics for both tasks. \emph{AADG} even generalizes well to challenging target domains, with DSC increases of 0.34\% on HRF (Domain B in Table~\ref{tab:vesselsota}), 2.59\% for OD and 3.03\% for OC on RIM-ONE-r3 (Domain B in Table~\ref{tab:odocsota}), while most of the other methods fail to generalize. We also conduct qualitative comparisons in Fig.~\ref{fig:rvs} and Fig.~\ref{fig:optic}. They clearly show that \emph{AADG} is more robust against different domains. Please note, the quantitative results on DoFE in Table~\ref{tab:vesselsota} and Table~\ref{tab:odocsota} are directly copied from the original paper, and therefore we do not include DoFE in our qualitative segmentation comparisons.

\begin{figure}[t]
\begin{center}
    \includegraphics[width=\linewidth]{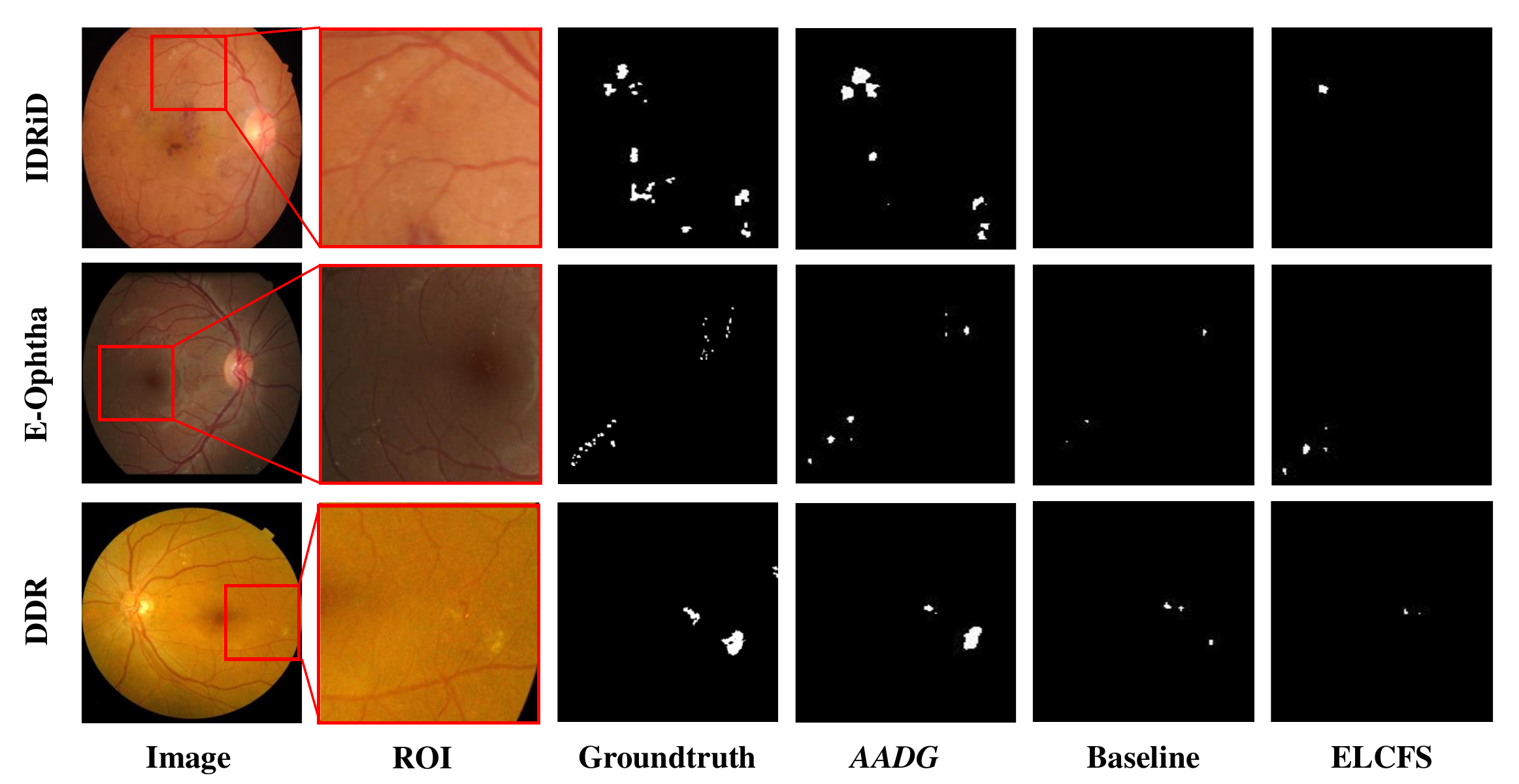}
\end{center}
   \caption{Qualitative comparisons of \emph{AADG}, the baseline and ELCFS on retinal lesion segmentation.}
\label{fig:hardex}
\end{figure}

\begin{figure}[t]
\begin{center}
    \includegraphics[width=\linewidth]{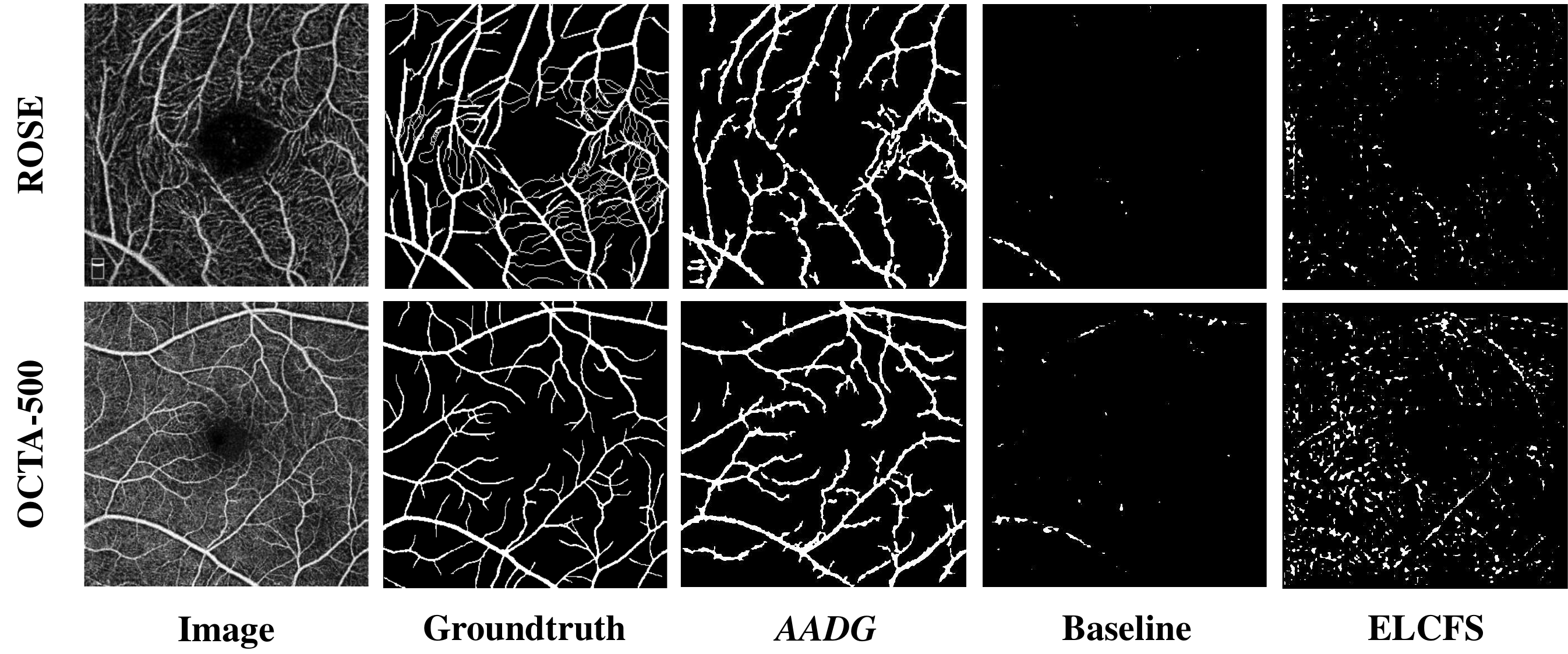}
\end{center}
   \caption{Qualitative comparisons of \emph{AADG}, the baseline and ELCFS on OCTA vessel segmentation.}
\label{fig:octa}
\end{figure}

\subsubsection{Comparison with DG methods on lesion segmentation} In the following two experiments, we only compare our \emph{AADG} with the baseline and ELCFS since ELCFS performs the best among all compared methods in our previous two tasks. As shown in Table~\ref{tab:hardex}, significant improvements from \emph{AADG} over ELCFS are observed; we obtain DSC increases of 2.32\%, 1.22\% and 2.31\% respectively on IDRiD, E-Ophtha and DDR. Representative segmentation results are presented in Fig.~\ref{fig:hardex}. HEs are common lesions in retinal fundus images, presenting as patches of irregular shapes and sizes. The intensity profile of HEs is close to the background, and it may vary depending on the global lighting conditions and imaging devices, causing potential performance degradations on different target domains. As shown in Fig.~\ref{fig:hardex}, with a strong domain shift, the baseline can barely detect HEs, whereas \emph{AADG} can correctly identify the major regions of the lesions. This indicates that our proposed method can effectively generalize models of interest to unseen domains.

\subsubsection{Cross-modality generalization on OCTA vessel segmentation} 
\begin{figure*}[thbp]
\begin{center}
    \includegraphics[width=0.8\textwidth]{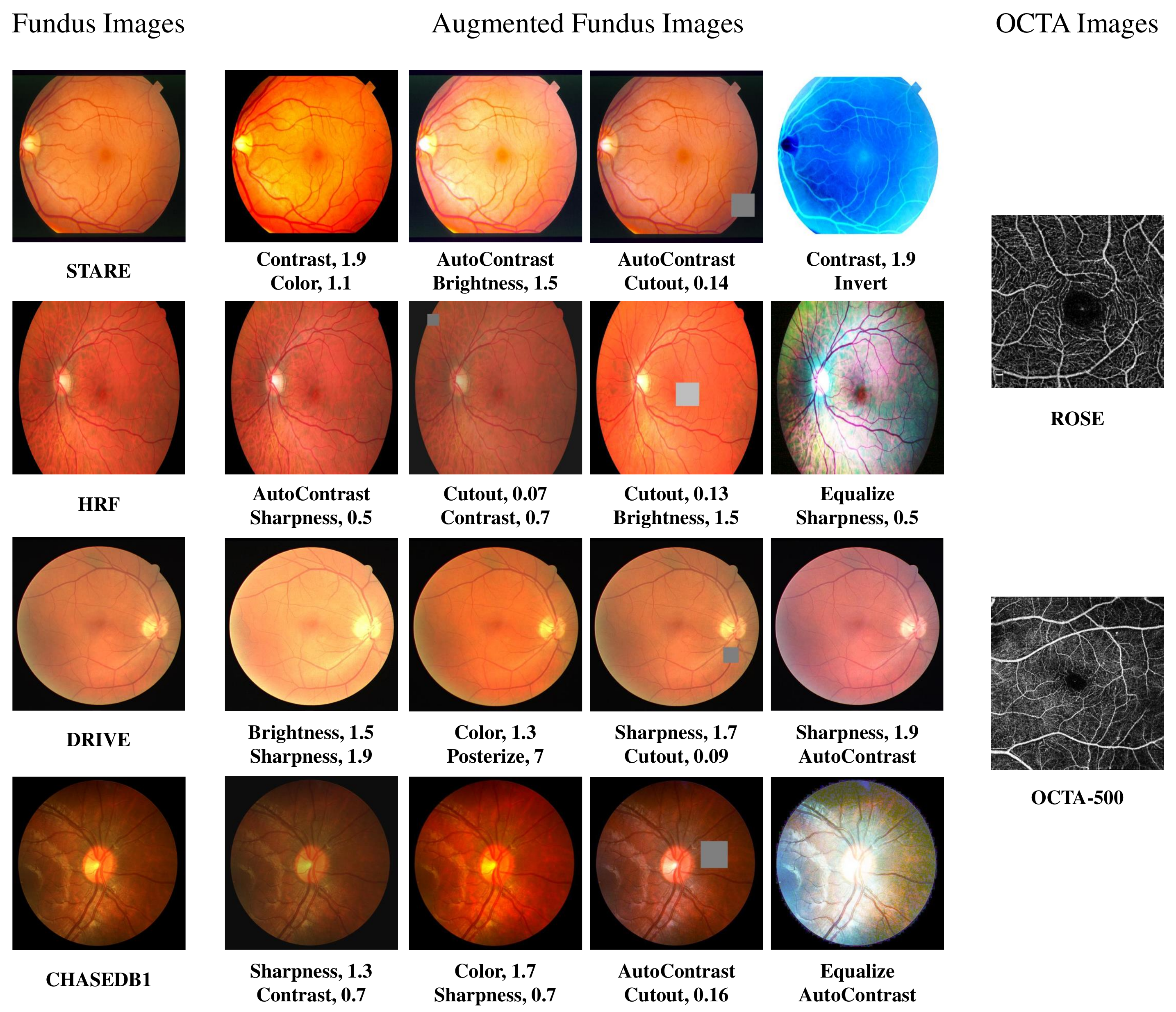}
\end{center}
   \caption{Visual assessment of representative augmented fundus images from \emph{AADG} in the fundus-to-OCTA cross-modality generalization experiment. The leftmost and rightmost columns respectively denote the source images and the target images, while the middle columns denote augmented source images under different transformation operations. The specific transformation operation and its corresponding magnitude are reported below each augmented image.}
\label{fig:augment}
\end{figure*}

To further validate the generalization ability of \emph{AADG}, we train \emph{AADG} on the DRIVE, STARE, HRF and CHASEDB1 datasets of fundus images and test it on the ROSE and OCTA-500 datasets of OCTA images. Both qualitative and quantitative results (Fig.~\ref{fig:octa} and Table~\ref{tab:octa}) show that \emph{AADG} can not only generalize well to images from different scanners but of the same modality, but also generalize well to images of different modalities. The compared ELCFS fails to segment the retinal vasculature from OCTA because it can only address the domain shifts among seen domains via frequency interpolation and cannot generate diversely and distinctly novel domains. As clearly shown in Fig.~\ref{fig:augment}, our learned policies not only meet various conditions of fundus images, but also accommodate the target domain data (OCTA images) very well (the rightmost one in the first row among the augmented fundus images).

\begin{table}[t]
\caption{Quantitative DSC comparisons of the baseline, BigAug, \emph{RADG} and \emph{AADG} on retinal vessel segmentation. Domains A, B, C, D are respective STARE, HRF, DRIVE and CHASEDB1. Paired Student's $t$-tests are conducted between \emph{AADG} and all other methods at a method level.}
\centering
\renewcommand{\arraystretch}{1.2}
\resizebox{0.95\columnwidth}{!}{
\begin{tabular}{l|llll|l|l}
\toprule
\multicolumn{1}{c|}{\textbf{Method}}    & \multicolumn{1}{c}{\textbf{A}} & \multicolumn{1}{c}{\textbf{B}} & \multicolumn{1}{c}{\textbf{C}} & \multicolumn{1}{c|}{\textbf{D}} &     \multicolumn{1}{c|}{\textbf{Average}}  & \multicolumn{1}{c}{\textbf{\textit{p}-value}}                       \\ \hline
Baseline & 76.32 & 72.23 & 76.27 & 75.71 & \multicolumn{1}{c|}{75.13} & \multicolumn{1}{c}{$<0.01$}\\
BigAug \cite{zhang2020generalizing} & 79.61 & 70.06 & 76.42 & 76.50 & \multicolumn{1}{c|}{75.65} & \multicolumn{1}{c}{$<0.01$}\\
\hline
\emph{RADG} & 81.22 & 71.32 & 75.75 & 77.83 & \multicolumn{1}{c|}{76.53} & \multicolumn{1}{c}{0.016}\\ 
\textbf{\emph{AADG}} & \textbf{81.79} & \textbf{72.57} & \textbf{77.70} & \textbf{78.34} & \multicolumn{1}{c|}{\textbf{77.60}} & \multicolumn{1}{c}{-} \\
\bottomrule
\end{tabular}
}
\label{tab:ablation}
\end{table}

\subsection{Ablation Study of AADG}
We conduct ablation analyses to investigate the mechanisms of \emph{AADG} in four folds: \textbf{1)} how does the augmentation search help to regularize the model, \textbf{2)} how does each transformation contribute to the generalization performance, \textbf{3)} what is an appropriate search space, and \textbf{4)} how is the computational efficiency.

\subsubsection{Contribution of diversity constraint} 
\begin{figure}[t]
\begin{center}
    \includegraphics[width=0.97\linewidth]{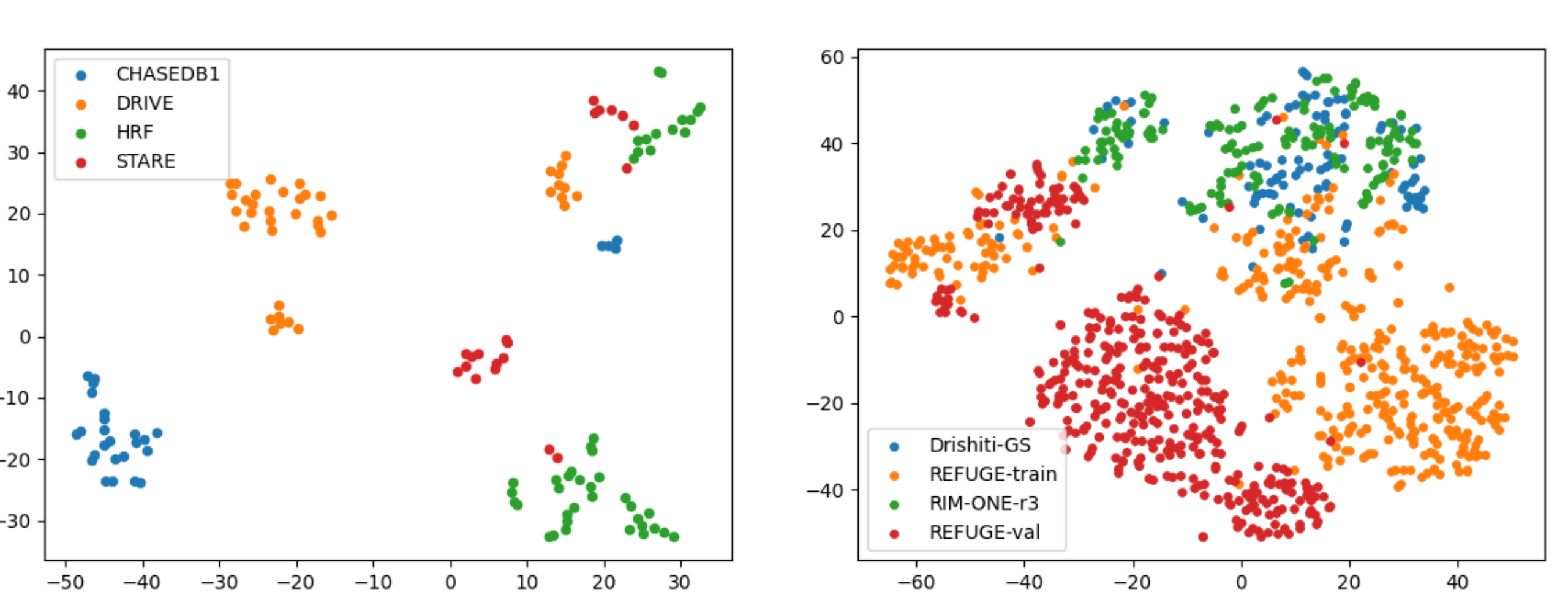}
\end{center}
   \caption{t-SNE visualization of VGG16 features of the augmented fundus images from \emph{AADG}. The left side presents the results of four datasets for vessel segmentation and the right side presents the four datasets for OD/OC segmentation. Different colors are used to denote different datasets.}
\label{fig:augtsne}
\end{figure}

\begin{figure}[t]
\begin{center}
    \includegraphics[width=0.98\linewidth]{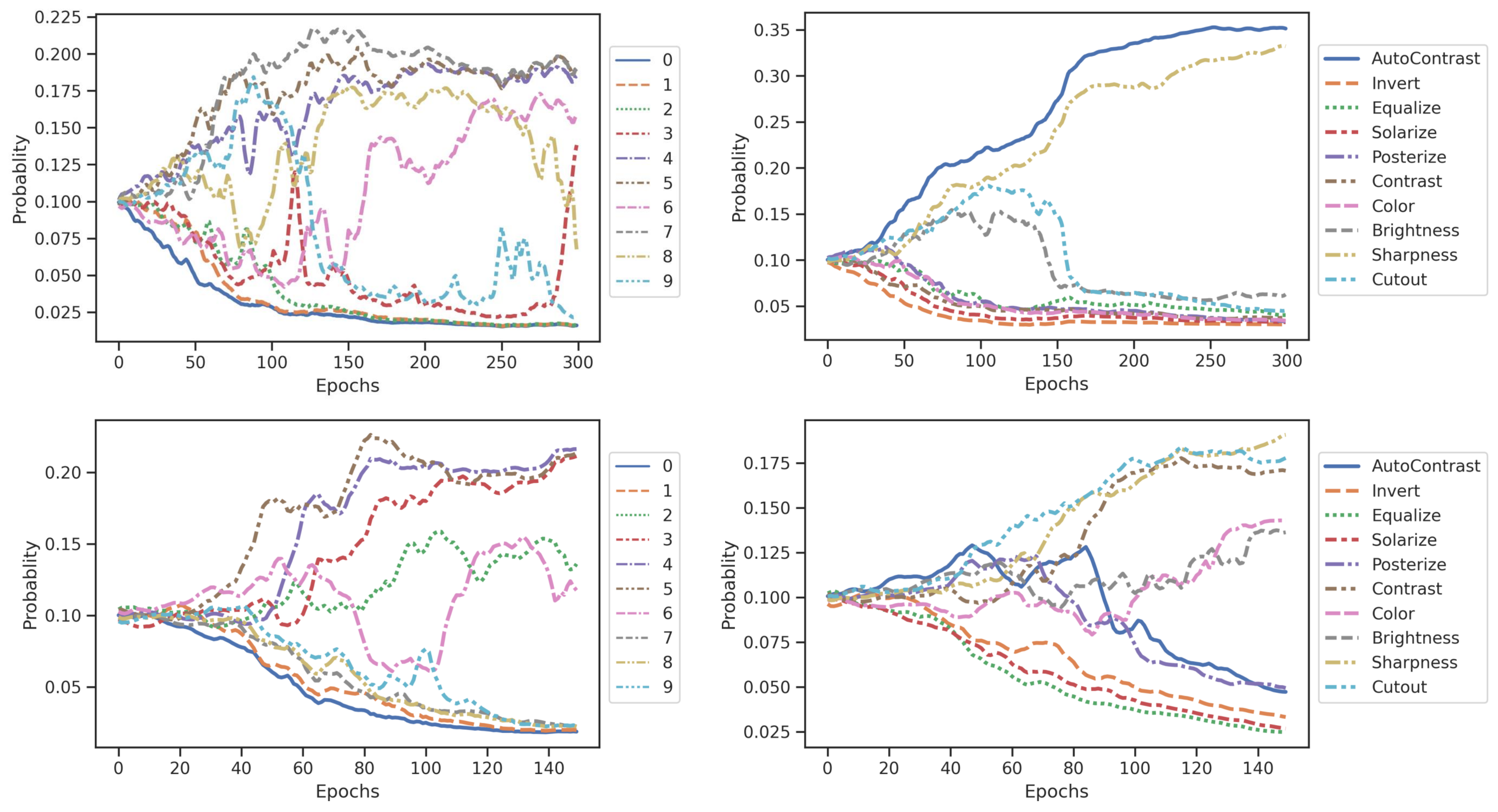}
\end{center}
   \caption{Probability distributions of the magnitudes and operations in the learned augmentation policies over epoch. The upper part of this figure shows the results from OD/OC segmentation and the lower part from vessel segmentation. We use different colors to denote different magnitudes and operations.}
\label{fig:tra}
\end{figure}

We first validate the effectiveness of the key component, namely the diversity constraint by excluding it from \emph{AADG}. Without the diversity constraint, \emph{AADG} is no longer an automated augmentation method. We name it as \emph{Random Augmentation for Domain Generalization (RADG)}, since in this case the controller uniformly samples augmentation policies from the search space. As shown in Table~\ref{tab:ablation}, \emph{RADG} improves the baseline by 1.40\% in DSC for retinal vessel segmentation. This result verifies our hypothesis that augmenting medical images with traditional image transformations can generate novel domains. \emph{RADG} even surpasses BigAug, which indicates that enumerating and permuting massive transformations do contribute to a large $C$, despite that many OOD data are also included. \emph{AADG}'s diversity constraint plays a vital role in enhancing the generalization performance with an average increase of 1.07\% in DSC, especially for HRF and DRIVE. The t-SNE visualization of the augmented images in Fig.~\ref{fig:augtsne} shows that the generated distributions are highly diverse. On the left side of Fig.~\ref{fig:tra}, we observe that \emph{AADG} avoids sampling the transformation with an extreme magnitude (e.g. $\lambda\in\{0,1,8,9\}$). The quantitative and qualitative results empirically reflect that maximizing the difference among the augmented novel domains can avoid OOD data and verify the weaker invariance assumption mentioned above. We perform statistical analysis by conducting paired Student's $t$-test between \emph{AADG} and other methods at a significance level of 0.05. In this experiment, we repeat each method for 10 times. As tabulated in Table~\ref{tab:ablation}, the differences between \emph{AADG} and all other methods are statistically significant. Collectively, we successfully demonstrate that automated augmentation is a practical and strong solution for DG.


\subsubsection{Contribution of each transformation} 
\begin{figure}[t]
\begin{center}
    \includegraphics[width=0.68\linewidth]{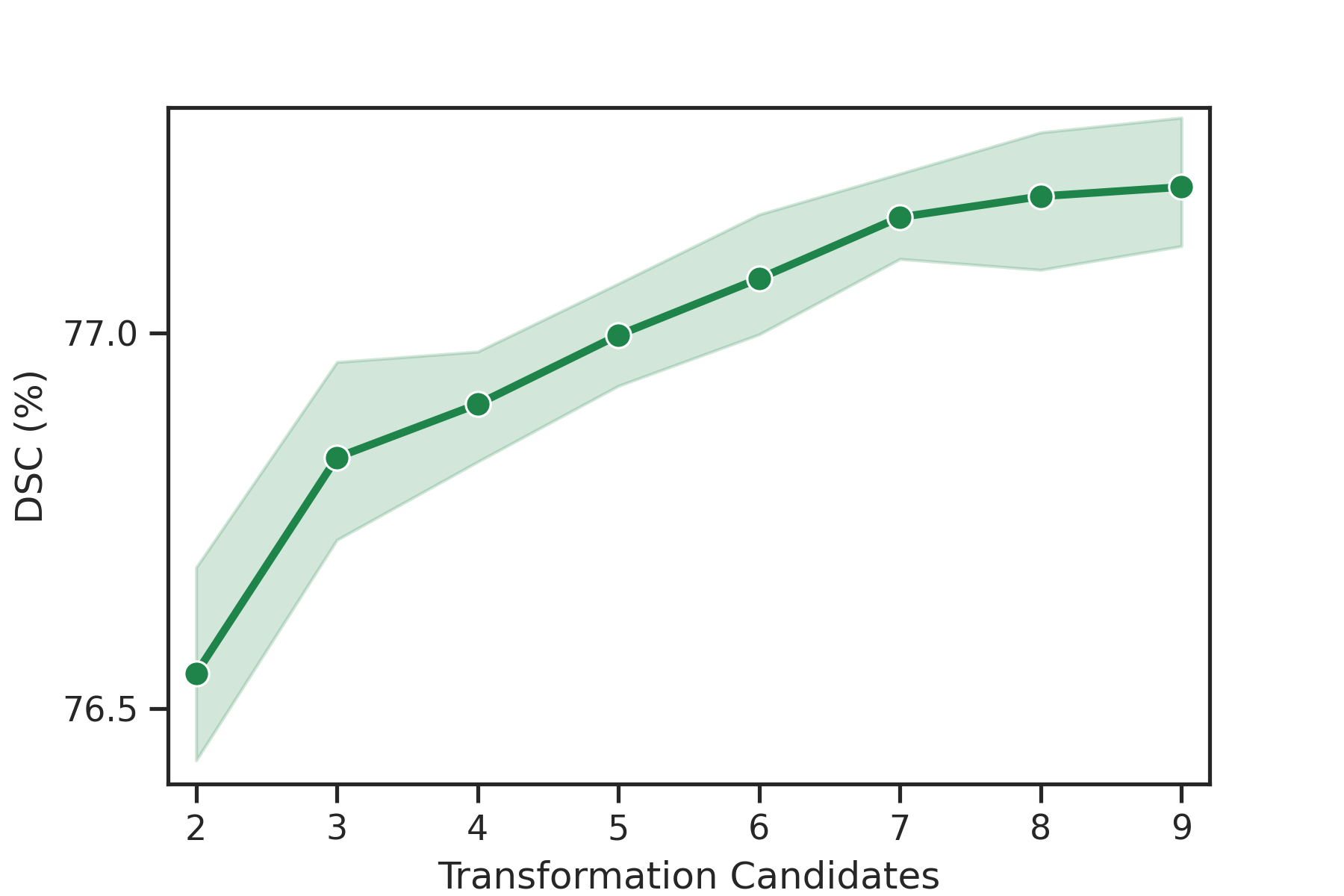}
\end{center}
   \caption{Results of varying the transformation candidates. On average, the performance improves when more transformation candidates are included in $\emph{AADG}$. The 95\% confidence intervals are plotted as the shaded region.}
\label{fig:trendex}
\end{figure}

We then investigate which transformation really works for DG. From the right side of Fig.~\ref{fig:tra}, we observe that the probabilities of the selected operations change over time. Considering OD/OC segmentation on REFUGE-val, AutoContrast and Sharpness are important for DG. For vessel segmentation on STARE, Sharpness, Cutout and Contrast matter. This indicates \emph{AADG} is explainable: we can easily identify the most important image transformations for a specific task and illustrate how the learned augmentation policy changes the images. Such results also confirm another motivation of this study that each dataset or task has its own best augmentation policy, even of the same image modality. 
The probability of a specific parameter does not monotonically increase or decrease, which means different operations and magnitudes play an essential role during the convergence of the segmentation model, establishing the effectiveness of our adversarial learning. Furthermore, we randomly sample subsets of transformations and train the \emph{AADG} framework for the vessel segmentation task. On one hand, Fig.~\ref{fig:trendex} suggests that the generalization performance of \emph{AADG} improves as the number of transformation candidates increases. Even with only two transformation candidates, which means \emph{AADG} only search for optimal magnitudes, \emph{AADG} improves the baseline by a considerable margin. This shows the importance of selecting an appropriate magnitude for each transformation. On the other hand, the monotonic increase reveals that all the transformation operations contribute to better generalization. To note, we repeat each experiment for 20 times and report the mean and 95\% confidence intervals.

\begin{figure}[t]
\begin{center}
    \includegraphics[width=0.98\linewidth]{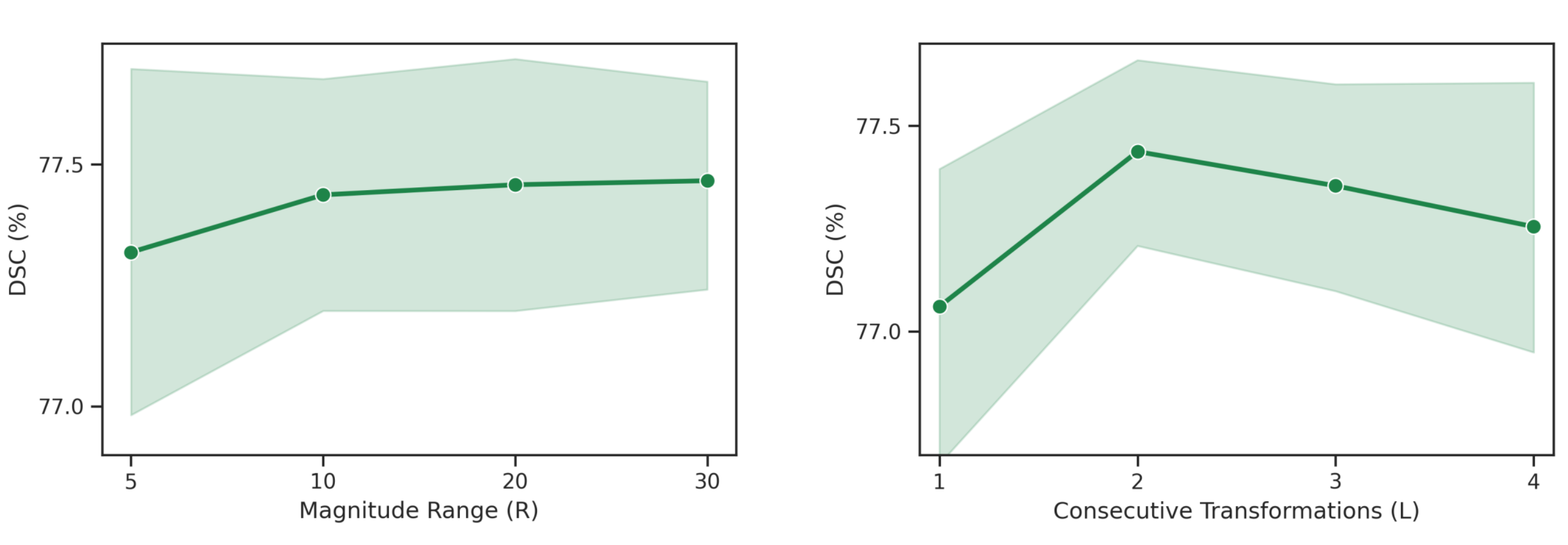}
\end{center}
   \caption{Results of varying $L$ and $R$.}
\label{fig:trend}
\end{figure}

\begin{table}[t]
\caption{Quantitative DSC comparisons between the baseline and Transfer, which is regularized by the learned augmentation policies from \emph{AADG} equipped with the backbone of MobileNetv2 for vessel segmentation.}
\centering
\renewcommand{\arraystretch}{1.2}
\resizebox{\columnwidth}{!}{
\begin{tabular}{ll|llll|l}
\toprule
\multicolumn{1}{c}{\textbf{Method}} & \multicolumn{1}{c|}{\textbf{Backbone}}    & \multicolumn{1}{c}{\textbf{A}} & \multicolumn{1}{c}{\textbf{B}} & \multicolumn{1}{c}{\textbf{C}} & \multicolumn{1}{c|}{\textbf{D}} &     \multicolumn{1}{c}{\textbf{Average}}                        \\ \hline
Baseline & \multicolumn{1}{c|}{EffNetv2-S} & 80.78 & 73.24 & 78.27 & 78.04 & \multicolumn{1}{c}{77.58} \\
Transfer & \multicolumn{1}{c|}{EffNetv2-S} & \textbf{82.43} & \textbf{73.95} & \textbf{78.84} & 
\textbf{79.35} & \multicolumn{1}{c}{\textbf{78.64}}\\
\hline
Baseline & \multicolumn{1}{c|}{SegFormer-B2} & 76.72 & 70.05 & 75.42 & 75.86 & \multicolumn{1}{c}{74.51} \\
Transfer & \multicolumn{1}{c|}{SegFormer-B2} & \textbf{81.97} & \textbf{71.07} & \textbf{77.64} & 
\textbf{77.94} & \multicolumn{1}{c}{\textbf{77.16}}\\
\bottomrule
\end{tabular}
}
\label{tab:transfer}
\end{table}

\subsubsection{Contribution of an appropriate search space} We further explore an appropriate search space to ensure computational efficiency and better convergence. We first try different magnitude ranges $R\in \{5,10,20,30\}$ and compare their generalization performance. A larger $R$ implies a more precise control of the magnitude $\lambda$, leading to less variation in the image intensity. The left side of Fig.~\ref{fig:trend} shows that the generalization performance slightly improves when $R$ increases from 5 to 10. A further increase of $R$ does not bring any more improvement, which means the performance is not sensitive to the choice of $R$. Thus we set $R=10$ in our entire study. We also evaluate the generalization performance of \emph{AADG} for different values of $L$, where $L\in \{1,2,3,4\}$. An increase in $L$ implies more combinations and permutations of the transformations of interest, involving more factors that cause domain shifts. This exponentially expands the search space. The right side of Fig.~\ref{fig:trend} clearly illustrates that a larger $L$ provides more diversity, at the cost of significantly increasing the optimization difficulty of DRL. As such, we set $L=2$ as the default value. 

\subsubsection{Computational Efficiency} The computational cost of \emph{AADG} during training is approximately $6\times$ more than that of the baseline. The training overhead comes from the reason that we sample $B=6$ policies to augment the image mini-batch. The training overhead caused by policy search is marginal, creditable to our small policy controller and the efficiency of the PPO algorithm. Nevertheless, we do not introduce any time overhead during inference since we do not add any layer to the segmentation model, which is one of the important advantages of \emph{AADG}.

\vspace{1mm}

\subsection{Transferability across Network Architectures} 
To further show the effectiveness of \emph{AADG}, the transferability of the augmentation policies is evaluated across different backbone networks. We first train \emph{AADG} equipped with the backbone of MobileNetv2, and then directly use the learned dynamic augmentation policies to regularize a stronger segmentation model on the union of all source datasets, named Transfer. Specifically, we train a DeepLabv3+ model with the backbone of EfficientNetv2-S (EffNetv2-S) \cite{tan2021efficientnetv2} and a model named SegFormer-B2 \cite{xie2021segformer} for vessel segmentation. Table~\ref{tab:transfer} presents that the learned policies can directly apply to other networks, improving over the baselines by 1.06\% and 2.65\% in DSC. This indicates our learned policies are model-agnostic. Once we have learned the policies for a small segmentation model, we can flexibly transfer them to any latest CNN-based or transformer-based SOTA model.


\section{Conclusion}
In this paper, we have proposed a novel data manipulation based DG method named \emph{AADG} for retinal fundus images. Specifically, we first formulated \emph{AADG} as an empirical risk minimization problem and established the search objectives. We then designed an appropriate search space and a corresponding novel proxy task with respect to the objectives. Adversarial learning and DRL were adopted to efficiently solve the problem. Comprehensive experiments on retinal vessel, OD/OC and lesion segmentation tasks successfully demonstrated the effectiveness and superiority of \emph{AADG} in multiple aspects. Future work will involve a more precise policy controller, sampling a policy for each image based on its characteristic to simultaneously address inter-domain and intra-domain discrepancy issues. Also, we will extend \emph{AADG} to more medical image analysis tasks such as disease classification and object detection and involve more regions of interest. Those directions may help reveal the rationale behind \emph{AADG} and better understand its learning behavior.

\vspace{1mm}

\bibliographystyle{IEEEtran}
\balance
\bibliography{ref.bib}

\end{document}